\documentclass[preprint,12pt,3p]{elsarticle}

\usepackage{amsmath,amssymb}
\usepackage{graphicx}
\usepackage{color}
\usepackage{subcaption}
\usepackage{siunitx}
\usepackage{dcolumn}
\usepackage{bm}
\usepackage{lineno}
\usepackage{tabularx}

\modulolinenumbers[5]
\begin{document}

\begin{frontmatter}

\title{A crossover of the solid substances solubility in supercritical fluids: what is it in fact?}

\author[isc]{Kalikin N.N. }
\author[isc]{Oparin R.D.}
\author[lpzg]{Kolesnikov A.L.}
\author[isc,hse]{Budkov Y.A. \corref{cor1}}
\author[isc]{Kiselev M.G.}
\address[isc]{G.A. Krestov Institute of Solution Chemistry of the Russian Academy of Sciences, Laboratory of NMR Spectroscopy and Numerical Investigations of Liquids, Akademicheskaya str. 1, 153045, Ivanovo, Russia}
\address[hse]{Tikhonov Moscow Institute of Electronics and Mathematics, School of Applied Mathematics, National Research University Higher School of Economics, 34, Tallinskaya Ulitsa, 123458, Moscow, Russia}
\address[lpzg]{Institut für Nichtklassische Chemie e.V., Permoserstr. 15, 04318, Leipzig, Germany}
\cortext[cor1]{Corresponding author}

\begin{abstract}
We investigate a well-known phenomenon of the appearance of the crossover points, corresponding to the intersections of the solubility isotherms of the solid compound in supercritical fluid. Opposed to the accepted understanding of the existence of two fixed crossover points, which confine the region of the inverse isobaric temperature dependence of the solubility, we have found that these points tend to shift with the change of the temperature and in the limit of the certain threshold value they converge to a single point. We demonstrate this analyzing the solubility data of a set of poorly soluble drug compounds, which have been computed in a wide area of the phase diagram via the approach, based on the classical density functional theory. Thorough analysis of the available in the literature experimental solubility data is found to be in an agreement with our conclusions, as one can find that the wider temperature region of the experimental study is, the more pronounced effect of the crossover points drift can be observed. 
\end{abstract}

\end{frontmatter}

There is a well-known phenomenon, observed when studying solid compounds' solubility in supercrtitical fluids, where one can locate the region of the so-called $"$retrograde vaporization$"$ (RV), where the increase of the temperature at constant pressure leads to the decrease of the studied compound's solubility \cite{chimowitz1986process,chimowitz1988analysis,foster1991significance}. The boundaries of this region are represented by two points, where all isotherms intersect and the solubility as a function of temperature has extremums, the corresponding pressure values are called the lower and upper crossover pressures \cite{esmaeilzadeh2008two,de2009solid}. Described phenomenon is rather interesting for practical use, since one can easily adjust the solubility via the change of the state parameters to achieve the appropriate conditions for a number of supercritical procedures, including micro- and nanonization, cocrystallization, precipitation, extraction, chromatography, etc. \cite{padrela2018supercritical,rodrigues2018pharmaceutical,gurikov2018amorphization,cocero2009encapsulation,kelley1990near,esmaeilzadeh2005supercritical}.

On the other hand, despite its practical importance and fundamental significance, the thorough investigation of the phenomenon of this area appearance is rather limited in literature \cite{chimowitz1988analysis,foster1991significance,esmaeilzadeh2008two,de2009solid}, where the occurrence of the crossover pressure was concluded to be a thermodynamic constraint in supercritical mixtures. It is also worth noting that the authors discussed only the upper crossover pressure, as they refer to it as a more important for the applications one. The variety of the experimental studies, devoted to the solubility measurements, also rarely demonstrate the location of the lower crossover pressure, partially for the same reason as mentioned above, but also due to the proximity of the lower point to the solvent critical point. 

Nevertheless the mentioned studies, describing the modeling of the position of the single point where all isotherms intersect, a careful investigation of the vast number of the available experimental solubility data for different chemical classes of the solutes \cite{wong1986solubilization,swaid1985nir,kramer1988solubility,baek2004density,pishnamazi2020measuring,wang2015measurement,knez2017solubility,gupta2006solubility}, especially in the cases of the studies demonstrating the broad temperature interval, leads one to the conclusion that either the accuracy of the experimental approaches and following correlation procedures leaves much to be desired, as the upper crossover point location often appears rather $"$smeared$"$, or that in actual fact there is no sole specific point where all isotherms intersect. The fact that the experimental solubility data, depicting the position of the lower crossover pressure point, to our knowledge, is practically absent in literature does not help one to grasp the idea, underlying this inconsistency.

In this communication, we share the results of our computational solubility study, which demonstrate that there are no two exact points in the concentration-pressure coordinates where all of the solubility isotherms intersect, allocating the region of the RV behavior. Instead, we observe that the points, corresponding to the temperature extremums of solubility, shift with the temperature change. Moreover, the increase of the temperature leads to the convergence of the lower and upper crossover pressure values to one point at certain temperature, after which the dependence of the solubility on temperature once again becomes direct for any pressure values.

The computation of the solubility isotherms was conducted with the help of the developed methodology, based on the solvation free energy calculation via the classical density functional theory (cDFT), the thorough description of which can be found elsewhere \cite{budkov2019possibility,kalikin2020carbamazepine}. The results of the methodology approbation have shown a satisfactory agreement with the available experimental data for a number of solutes. Here we have obtained data for six poorly-soluble drug compounds of different chemical structure with analgesic, anti-inflammatory, anticonvulsive, antioxidant and neuroprotective properties \cite{todd1990naproxen,rainsford2009ibuprofen,cotton1985diflunisal,ballenger1978therapeutic,awtry2000aspirin,surov2015impact}, namely diflunisal, ibuprofen, naproxen, aspirin, carbamazepine and thiadiazole derivative (1-[5-(3-chloro-4-methyl-phenylamino)-1,2,4-thiadiazol-3-yl]-propan-2-ol). For each compound we have obtained a set of isotherms, corresponding to the temperatures from 313.15 K to 483.15 K with a step of 10 K, and several auxiliary ones to differentiate the areas where the region of the phenomenological behavior is supposed to vanish. The pressure increment of 5 bar and region from 75 bar to 330 bar allowed us to determine the position of both the upper and lower crossover region boundaries for each solute. We should note that obtained solubility values for each drug compound are in a satisfactory agreement with the available in the literature experimental data, and with the results of the solubility measurements, based on our infrared spectroscopy approach, for the thiadiazole derivative, for which there are no available in literature solubility data (see Appendix for the details of the experimental approach and the thorough comparison).

\begin{figure}[h!]
\center{\includegraphics[width=1\linewidth]{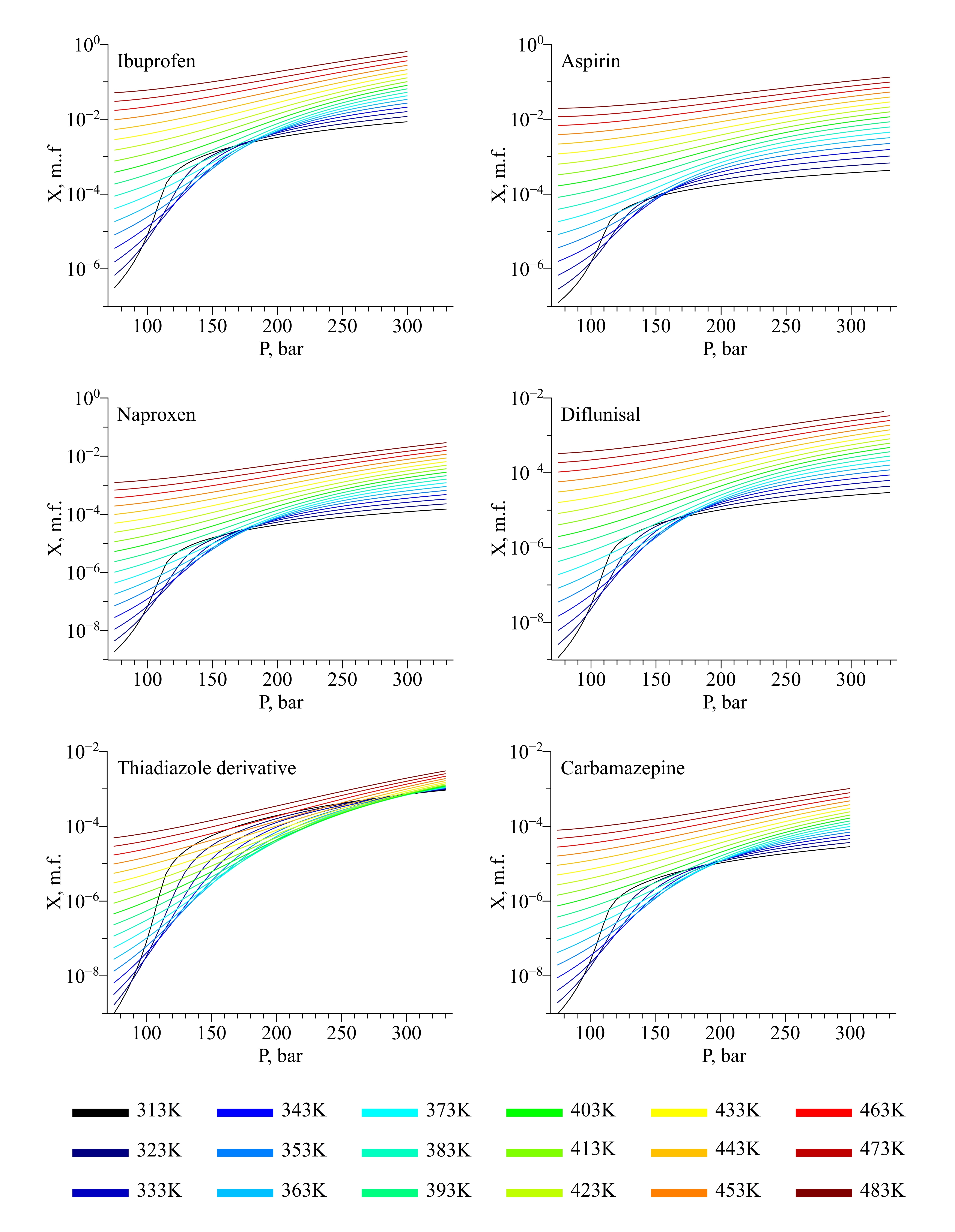}}
\caption{Solubility isotherms of each compound.}
\label{Fig1}
\end{figure}

Turning to the obtained results, at Fig. \ref{Fig1} we present the computed solubility isotherms for each compound. As it is clearly seen, there are no two specific crossover points where the temperature derivatives of all solubility isotherms simultaneously change the sign. Rather one can see that the location of such points shifts with the temperature increase, which is particularly apparent for the lower crossover point. Moreover, the pressure values of the first lower crossover points for all compounds is similarly located in the vicinity of the fluid's critical pressure, whereas the first upper crossover points appearance varies in pressure from approximately 150 bar for aspirin to 290 bar for thiadiazole derivative. It is also natural that the crossover region disappears, starting from a certain for each compound temperature value. Nevertheless, the solubility values are qualitatively reasonable even for high temperature values.

\begin{figure}[h!]
\center{\includegraphics[width=1\linewidth]{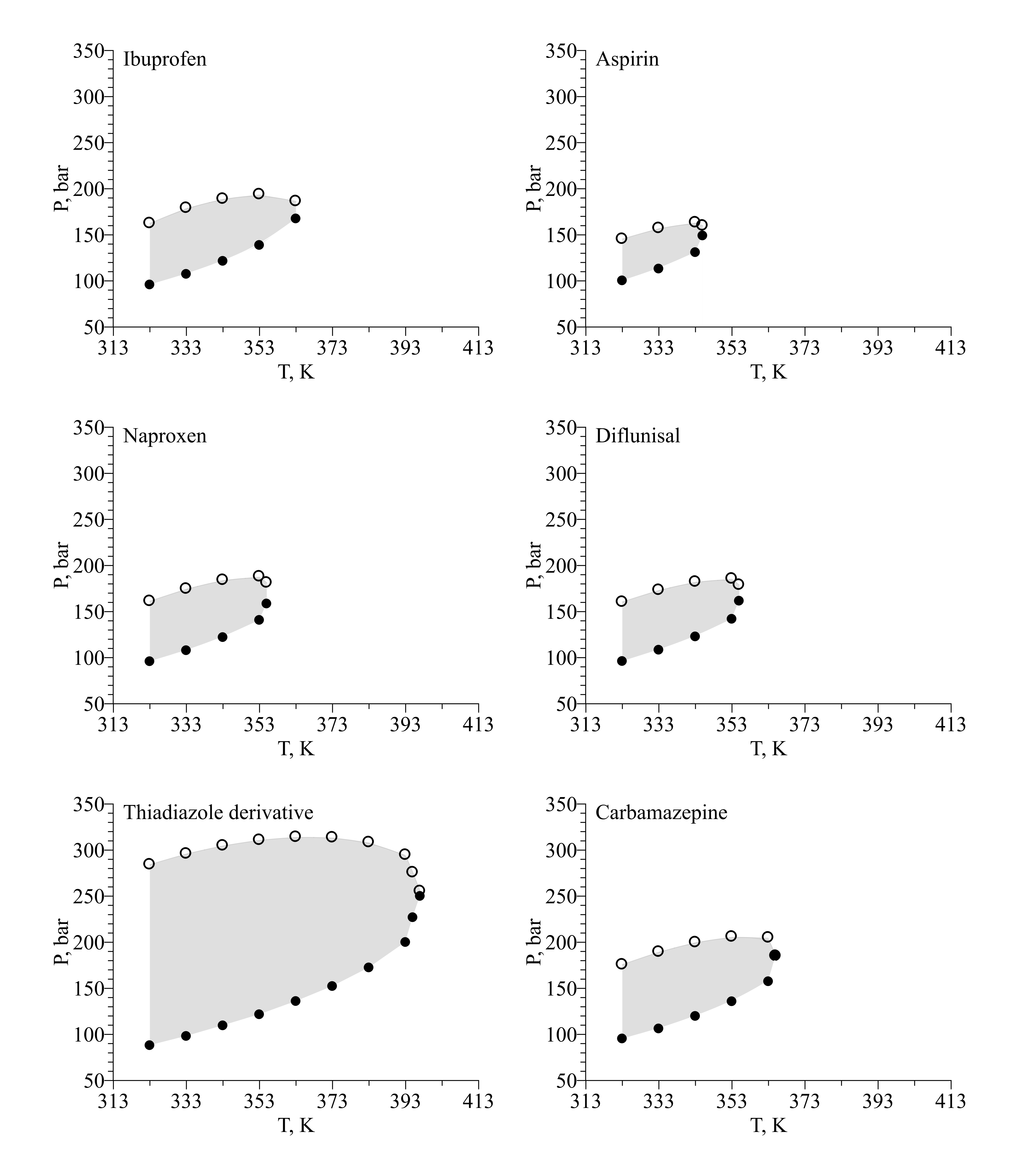}}
\caption{Change of the crossover points positions, bounding the area (in grey) of the inverse isobaric dependence of the solubility on temperature, i.e. where $\partial X/\partial T<0$, for each studied solute. The filled and open circles  depict the lower and upper crossover points correspondingly. The boundary of the grey area, consisting from the crossover points, corresponds to the condition of the $\partial X/\partial T=0$.}
\label{Fig2}
\end{figure}

Fig. \ref{Fig2} shows the RV region of each compound in the P-T coordinates, namely the grey-filled area is the region, where the solubility of the compound decreases with the temperature increase at constant pressure, i.e. where $\partial X/\partial T<0$. The boundary of this area consists of the lower and upper crossover $"$lines$"$, composed of the upper (open circles) and lower (filled circles) crossover points, i.e. points where $(\partial X/\partial T)_P=0$. As the temperature increases they converge to a single point at the certain threshold temperature value, above which $\partial X/\partial T>0$ for any pressure values. It is worth noting that the behavior of the crossover lines is similar for all compounds, namely the lower crossover line monotonically increases with the temperature increase, whereas the upper one reveals an unpronounced maximum. It is clear that the observed shape of the crossover region is drastically different from the rectangular one, which one can expect to see, when considering the $"$classical$"$ case of two constant crossover points. In the latter case it is also unclear, if such representation of the crossover region is valid in the limit of high temperatures, as in contrast to our results, there is no self-evident boundary at some temperature value.

\begin{figure}[h!]
\center{\includegraphics[width=1\linewidth]{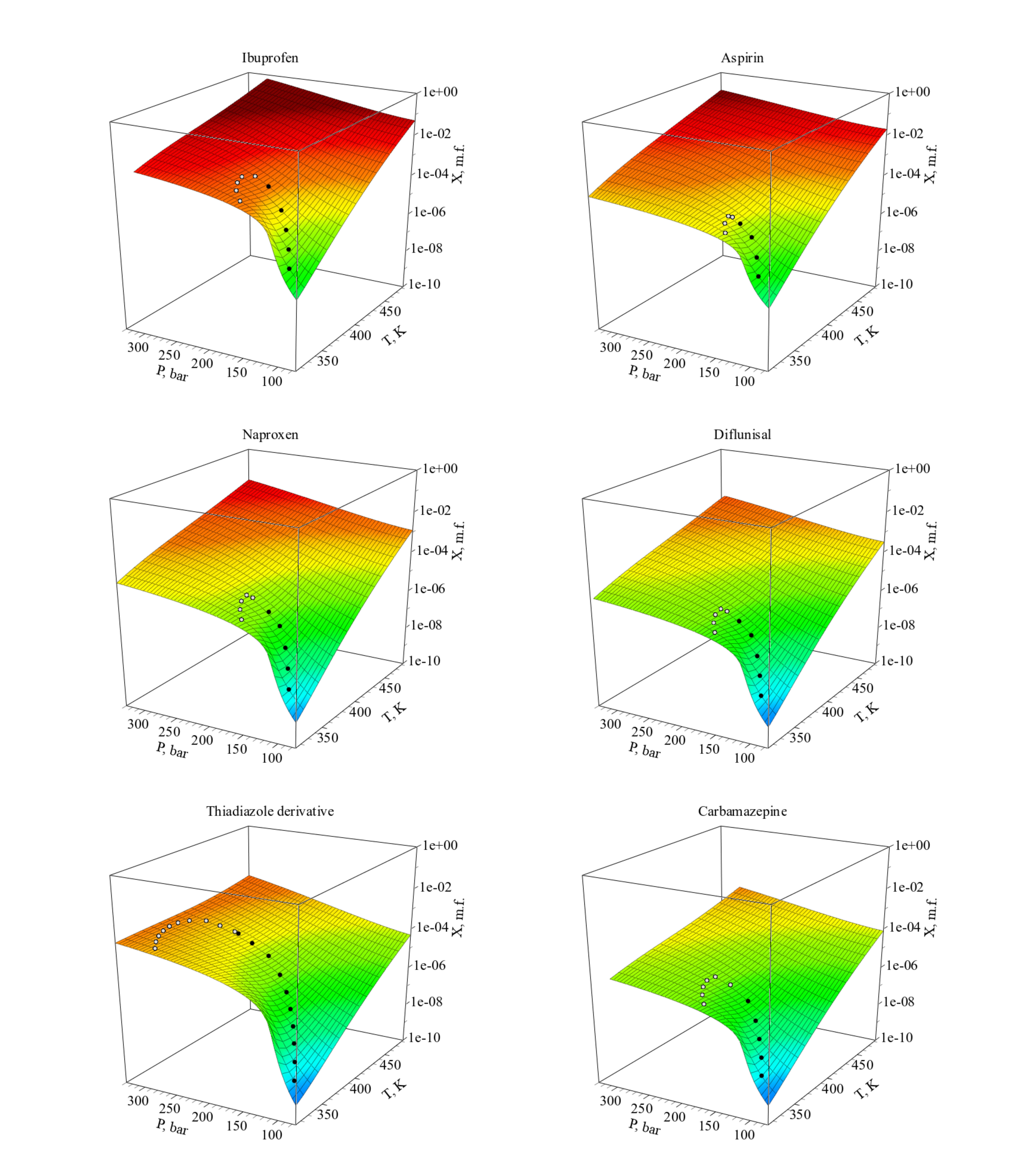}}
\caption{Solubility surface for each drug compound. Filled and open circles correspond to the lower and upper crossover points, respectively, i.e. where $\partial X/\partial T=0$.}
\label{Fig3}
\end{figure}

Finally, at the Fig. \ref{Fig3} we show the solubility surfaces of the studied compounds, plotted from the obtained solubility data, on which we have outlined the crossover regions from the Fig. \ref{Fig2}. Looking at such surface, one can easily specify the location of the crossover region and identify the phenomenological change of the solubility values for each compound. It is also worth noting that obtaining such surfaces, basing on the experimental techniques or fully atomic computer simulations is a nontrivial task regarding the time and resources costs.

As a discussion, we would like to underline that the behavior we have observed does not contradict the existing experimental solubility data. We suppose that the drift of the upper crossover point, obvious from the computational results, is rarely observed in the experimentally measured data mostly due to the fact that the solubility is in general measured in the narrow temperature interval. As it can be seen from the results of our calculations at Fig. \ref{Fig2} the rate of the upper crossover pressure position change is lower than that of the lower one. Thus, in the case of the narrow temperature interval the change in the upper crossover point position during the experimental investigations can be overlooked due to the experimental error or the inaccuracy of the correlation procedure. On the other hand, the change in the lower crossover pressure position should be more pronounced even in the narrower temperature interval. As it was mentioned above, the experimental solubility data, demonstrating the lower crossover point, is rather hard to find. Nevertheless, the ref.\cite{gregorowicz2003solubility} demonstrates the solubility of the eicosane in supercritical ethane and ethylene in a rather wide range of temperatures of 330.15--370.15 K and 287.15--348.15 K, respectively, with the appearance of the same trend of the lower crossover point shift as in our results. These results also demonstrate that the observed behavior is not a specific feature of the scCO$_2$ solvent. Although there are data for several solutes in different supercritical fluids (see ref.\cite{mendes1999solubility,schmitt1986solubility,singh1993solubility,kurnik1981solubility} and data within the ref.\cite{foster1991significance}), where the upper crossover pressure point, located via the experimental data correlation procedures, has a concrete position, the temperature ranges, studied in these papers, did not exceed 30 K, which can be concluded to be too narrow to observe a pronounced effect. From the ref.\cite{tsai2014solubility}, reporting on the solubility of the niflumic acid and celecoxib for the temperature ranges of 313.15--353.15 K and 323.15--343.15 K, respectively, it can be seen that the more broad the temperature interval is, the more pronounced the effect of the crossover point shift. Namely, the experimental points and correlation curves demonstrated more accurate intersection of the isotherms in the case of the celecoxib, while for the niflumic acid one can not observe a single crossover point. The disappearance of the upper crossover pressure with the temperature growth can be clearly observed from the set of the experimental studies using flame ionisation detector method in a wide temperature range \cite{miller1995determination,miller1996solubility,miller1997solubility}, when comparing with the experimental solubility investigations of the same solutes, but in the narrower temperature interval \cite{li2003solubility,lou1997temperature,bartle1990measurement}. Moreover, although accurate up-to-date experimental solubility data of the drug compounds \cite{pishnamazi2020using,shojaee2013experimental,pishnamazi2020thermodynamic,pishnamazi2020measuring,zabihi2020loxoprofen,zabihi2021measuring,sodeifian2019solubility,sodeifian2017determination} is rarely investigated in the wide temperature range, the "smearing" of the crossover point is obvious when plotting solubility data in the logarithmic scale in comparison with the linear scale.

While conducting a solubility computation for a set of the solutes in a wide area of the phase space, we observed the divergence with the conventional understanding of the phenomenon of the crossover region appearance in the supercritical fluid mixtures. We have found that there are no two specific crossover points at the phase diagram region where all of the solubility isotherms intersect, rather such characteristic points, where the temperature derivative of the solubility equals to zero, shift with the change of the temperature, constructing two lines. In the limit of the certain threshold temperature value, they converge to a single point, after which the isobaric inverse dependence of the solubility on temperature changes to the direct one for any pressure value. Our observation is also supported by rather clear evidence of the discussed behavior in the available experimental solubility data.  

\section*{Appendix}

\subsection*{Solubility of thiadiazole derivative (1-[5-(3-chloro-4-methyl-phenylamino)-1,2,4-thiadiazol-3-yl]-propan-2-ol) in scCO$_2$}
\begin{figure}[h!]
\center{\includegraphics[width=0.7\linewidth]{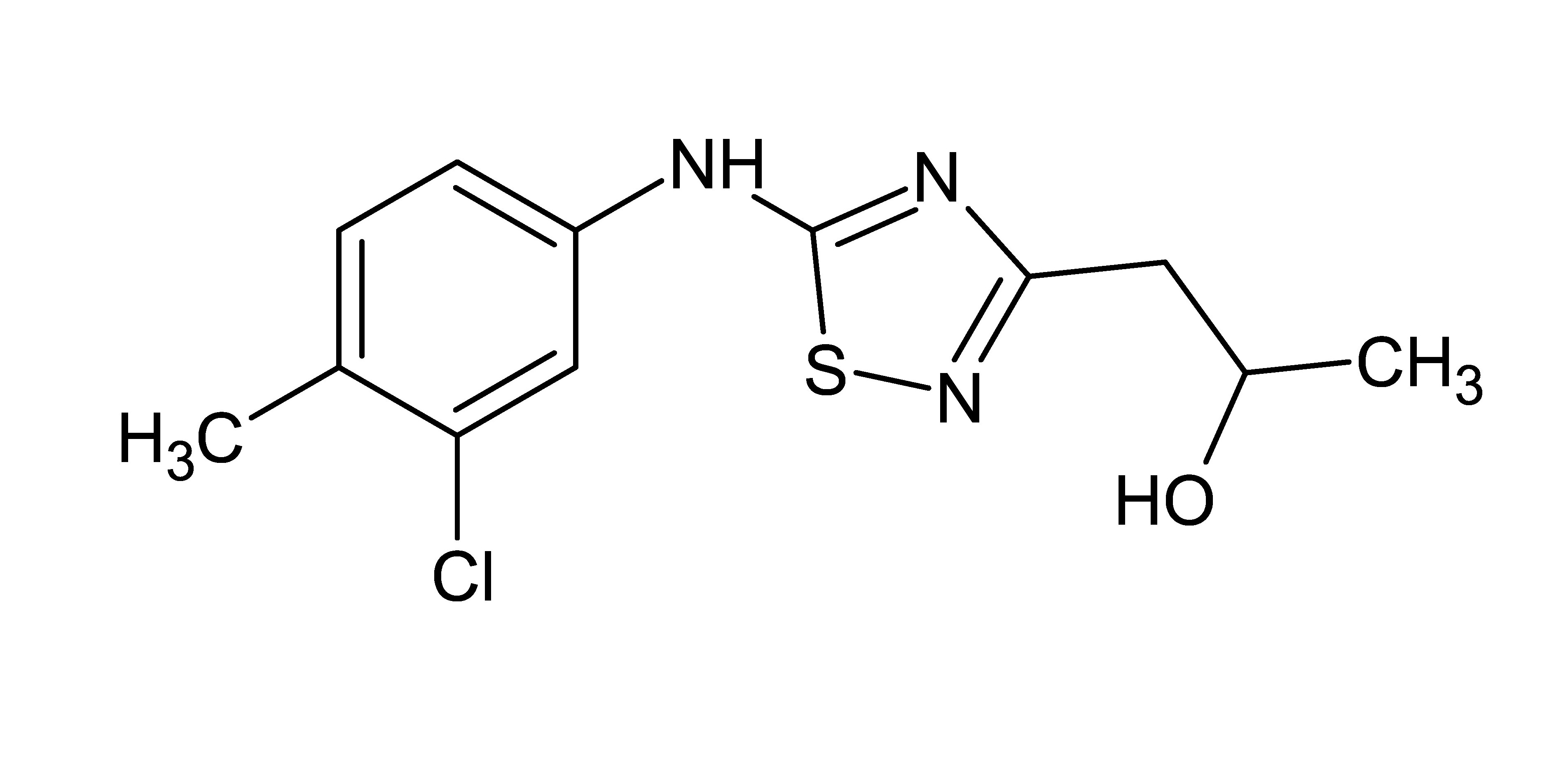}}
\caption{Molecular structure of the studied thiadiazole derivative.}
\label{Fig01supp}
\end{figure}

In this work we measured the solubility (concentration of a saturated solution) of thiadiazole derivative (structural formula shown at Fig. \ref{Fig01supp}) in scCO$_2$ as a function of temperature in the temperature range of 313.15--393.15 K along the isochore, corresponding to CO$_2$ density equal to 1.3 of its critical value ($\rho_{cr}=10.6249$ mol/l). Here we used a self-consistent approach developed by us (see e.g. refs. \cite{oparin2016new,kalikin2020carbamazepine}). Within this approach, based on the Beer-Lambert law, we use the integral extinction coefficient $\varepsilon_{int}$ (molar absorption coefficient) value of a chosen analytical spectral band.

\begin{figure}[h!]
\center{\includegraphics[width=0.7\linewidth]{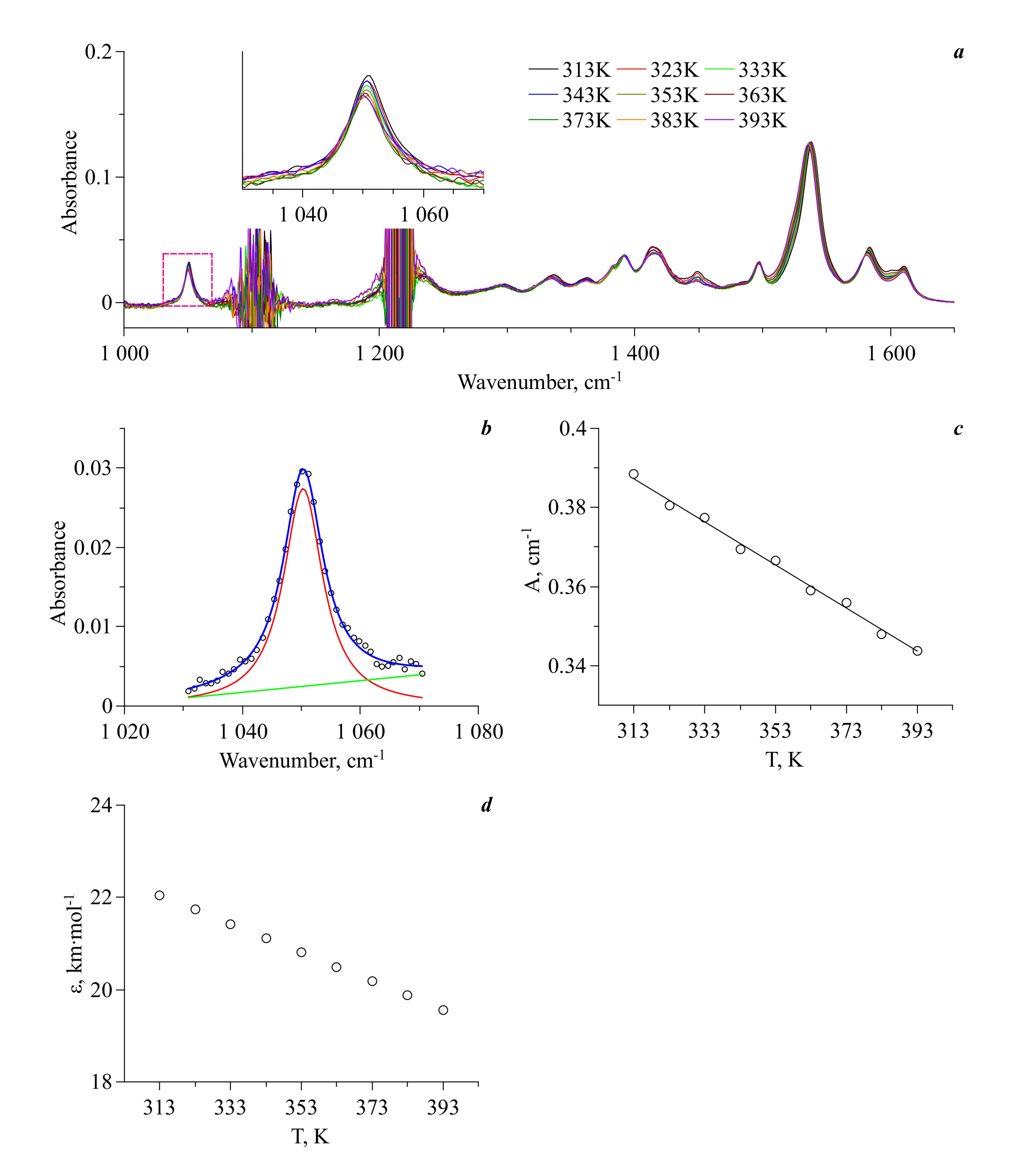}}
\caption{a. Experimental IR spectra of thiadiazole derivative dissolved in CHCl$_3$ (solute concentration 1.2549$\cdot 10^{-2}$ mol/l), area limited with rectangular corresponds to the analytical spectral band, insert corresponds to the enlarges analytic spectral band area); b. Example of the fitting of the analytical spectral band at $T=383.15$ K (dots -- experimental spectrum, blue line -- superposition, red line -- modeling spectral band, green line -- base line); c. Temperature dependence of the modeling spectral band integral intensity, obtained from the fitting procedure (dots -- data, line -- fitting line); d. Temperature dependence of the calculated extinction coefficient of analytical spectral band.}
\label{Fig1supp}
\end{figure}

On the first stage, in order to define the temperature dependence of $\varepsilon_{int}$ we measured the IR spectra of thiadiazole derivative in its solution in chloroform (CHCl$_3$) with concentration of the solute ($c$) equal to $1.2549\cdot10^{-2}$ mol/l. The spectra were measured on FT-IR spectrometer Bruker Vertex 80v using special high pressure high temperature (HPHT) optical cell with a constant volume developed by us. This cell as well as the experimental setup are described in details in our previous works (see e.g. refs. \cite{oparin2014dynamic,oparin2019polymorphism}). The spectra were measured in the wavenumber range of 1000--4000 cm$^{-1}$ with a resolution of 1 cm$^{-1}$, the optical path length ($l$) was 0.140 mm. Set of the experimental spectra in the wavenumber range of 1000--1650 cm$^{-1}$ is presented in Fig. \ref{Fig1supp}a. The spectral band in the domain of 1030--1070 cm$^{-1}$ that is related to deformation vibrations of Cl - methyl - substituted benzene ring in molecule of thiadiazole derivative (see Fig. \ref{Fig01supp}) was chosen as analytical. To calculate the integral intensity of this band ($A$) we applied standard procedure of spectra fitting based on the non-linear curve approximation using Fityk software package \cite{wojdyr2010fityk}. Example of the fitting of the analytical spectral band is shown in the Fig. \ref{Fig1supp}b. The dependence of $A=f(T)$ presented in Fig. \ref{Fig1supp}c can be described by a linear equation with high accuracy, confirming the correctness of the analytical spectral band choice (see e.g. refs. \cite{oparin2016new,kalikin2020carbamazepine,oparin2014dynamic}). Then, the data obtained by the linear fit were used to calculate temperature dependence of $\varepsilon_{int}$ according to following equation:

\begin{equation}
\varepsilon_{int}=\frac{A(T)}{l\cdot c}
\end{equation}

The dependence of $\varepsilon_{int}$ is presented in Fig. \ref{Fig1supp}d. All these values as well as the values of the intensity and data obtained by their linear fit are tabulated in Table \ref{table_supp_1}.

\begin{figure}[h!]
\center{\includegraphics[width=0.7\linewidth]{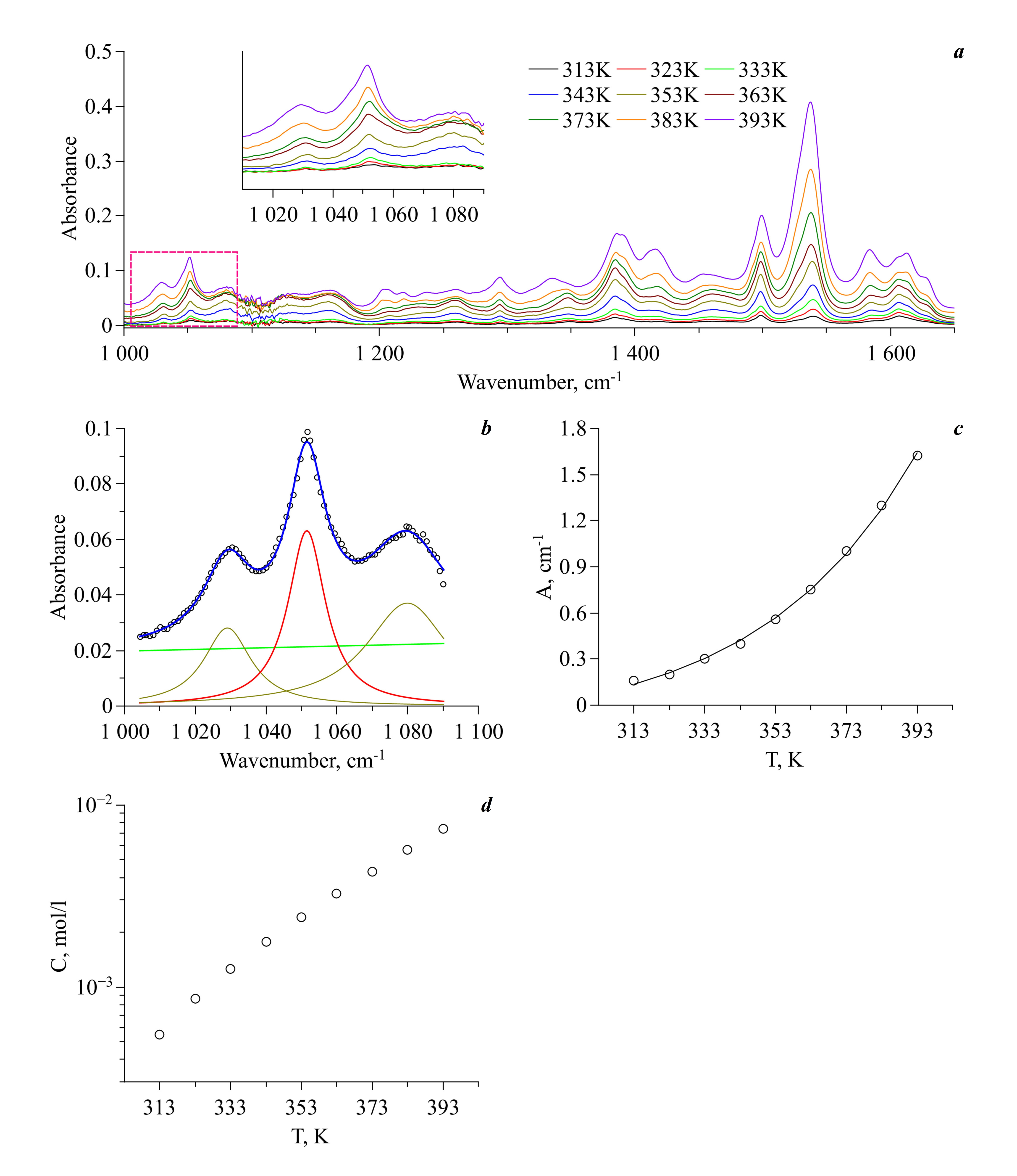}}
\caption{a. Experimental IR spectra of saturated solution of thiadiazole derivative in scCO2, area limited with rectangular corresponds to the analytical spectral region, insert corresponds to the enlarges analytic spectral band are); b. Sample of the fitting of the analytical spectral band at $T=383.15$ K (dots -- experimental spectrum, blue line -- superposition, red line -- spectral profile corresponding to analytical spectral band (see Fig. \ref{Fig1supp}a), dark yellow lines -- other spectral contributions, green line -- base line); c. Temperature dependence of the integral intensity of the modeling spectral profile corresponding to the analytical spectral band (dots -- data, line -- fitting line). d. Temperature dependence of the calculated concentration of saturated solution of thiadiazole derivative in scCO$_2$.}
\label{Fig2supp}
\end{figure}

In order to define the concentration of a saturated solution of thiadiazole derivative in scCO$_2$, on the second stage we measured the IR spectra of thiadizazole derivative in its saturated solution in scCO$_2$ being in permanent contact with the excess of the crystalline thiadiazole derivative in the same temperature range along the same isochore. For these measurements we used the same HPHT cell but with the optical path length of 1.127 mm. The spectra were also measured in the wavenumbers range of 1000--4000 cm$^{-1}$ with a resolution of 1 cm$^{-1}$ and presented in Fig. \ref{Fig2supp}a. We used the same procedure of spectra fitting to define the temperature dependence of the integral intensity of the analytical spectral band (see Fig. \ref{Fig2supp}b). The dependence of $A=f(T)$ is presented in Fig. \ref{Fig2supp}c. As would be expected, this dependence can be fitted with the high accuracy by the exponential equation, which is typical for such slightly soluble in scCO$_2$ organic substances (see e.g. refs. \cite{oparin2016new,kalikin2020carbamazepine,oparin2014dynamic}).

Then, we used these data to calculate the concentration of a saturated solution of the thiadiazole derivative in scCO$_2$ solving the inverse task:

\begin{equation}
    c(T)=\frac{A(T)}{l\cdot\varepsilon_{int}}
\end{equation}

The dependence of the thiadiazole derivative solubility in scCO$_2$ ($c=f(T)$) is presented in Fig. \ref{Fig2supp}d. We also tabulated these values along with the values of $A$ and data obtained by their exponential fit (see Table \ref{table_supp_1}). In this table we also presented the molar fraction values ($X$) of the thiadiazole derivative in scCO$_2$ as a function of temperature. $X$ values were calculated following next equation:

\begin{equation}
    X=\frac{c_{solute}}{c_{solute}+\rho_{solvent}},
\end{equation}
where $\rho_{solvent}$ is CO$_2$ density equal to 1.3$\cdot\rho_{cr}$.

\begin{table}[h!]
\centering
\caption{Extinction coefficient ($\varepsilon_{int}$) of the analytical spectral band calculated on the basis of integral intensity $A^*$, $c$ -- concentration of a saturated solution of thiadiazole derivative in scCO$_2$ calculated on the basis of the integral intensity $A^{**}$ and ($\varepsilon_{int}$), $X$ -- molar fraction of thiadiazole derivative in the solution corresponding to the concentration $c$. $A$ -- integral intensity of analytical spectral band, $A^*$ -- integral intensity of analytical spectral band fitted by linear equation, $A^{**}$ -- integral intensity of analytical spectral band fitted by the exponential equation.}
\begin{tabular}{c|c|c|c|c|c|c|c}
\footnotesize
& \multicolumn{3}{c}{Thiadiazole derivative in CHCl$_3$} & \multicolumn{4}{|c}{Thiadiazole derivative in scCO$_2$} \\
      \hline
       $T$,K & $A$,[cm$^{-1}$]  & $A^*$,cm$^{-1}$ & $\varepsilon_{int}$,cm$\cdot$ mol$^{-1}$ & $A$,cm$^{-1}$ & $A^{**}$,cm$^{-1}$ & $c$,mol/l &  $X$,m.f. \\
      \hline
313.15	&	0.389	&	0.3872	&	2.2041$\cdot10^6$	&	0.161	&	0.1357	&	5.4644$\cdot10^{-4}$	&	3.9561$\cdot10^{-5}$	\\
323.15	&	0.381	&	0.3818	&	2.1731$\cdot10^6$	&	0.200	&	0.2101	&	8.5771$\cdot10^{-4}$	&	6.2095$\cdot10^{-5}$	\\
333.15	&	0.378	&	0.3763	&	2.1422$\cdot10^6$	&	0.301	&	0.3034	&	1.2566$\cdot10^{-3}$	&	9.0972$\cdot10^{-5}$	\\
343.15	&	0.370	&	0.3709	&	2.1112$\cdot10^6$	&	0.396	&	0.4205	&	1.7674$\cdot10^{-3}$	&	1.2795$\cdot10^{-4}$	\\
353.15	&	0.367	&	0.3655	&	2.0802$\cdot10^6$	&	0.560	&	0.5676	&	2.4211$\cdot10^{-3}$	&	1.7526$\cdot10^{-4}$	\\
363.15	&	0.359	&	0.3600	&	2.0492$\cdot10^6$	&	0.750	&	0.7522	&	3.2572$\cdot10^{-3}$	&	2.3577$\cdot10^{-4}$	\\
373.15	&	0.356	&	0.3546	&	2.0182$\cdot10^6$	&	1.000	&	0.9840	&	4.3263$\cdot10^{-3}$	&	3.1313$\cdot10^{-4}$	\\
383.15	&	0.348	&	0.3491	&	1.9872$\cdot10^6$	&	1.300	&	1.2750	&	5.6932$\cdot10^{-3}$	&	4.1202$\cdot10^{-4}$	\\
393.15	&	0.344	&	0.3437	&	1.9562$\cdot10^6$	&	1.621	&	1.6404	&	7.4406$\cdot10^{-3}$	&	5.3841$\cdot10^{-4}$	\\
 \end{tabular}
 \label{table_supp_1}
\end{table}

\subsection*{Comparison of experimental solubility data with that obtained using DFT calculation.}
Computational solubility approach is based on the equilibrium condition between the solution phase and solute's solid phase
\begin{equation}
X\approx \frac{p^{sat}}{\rho_b k_BT}\exp(\beta\nu^s[p-p^{sat}]-\beta\Delta G_{solv}).
\label{slblt}
\end{equation}
In the framework of the proposed methodology we extract the solute's vapor pressure $p^{sat}$ from the experimental literature data, molar volume of the solute $\nu^s$ is determined on the basis of the group contribution methods \cite{immirzi1977prediction,cao2008use} and the solvation free energy $\Delta G_{solv}$ we compute basing on the classical density functional theory (cDFT), where particles of the solute and solvent are modeled as coarse-grained hard spheres, interacting through the effective Lennard-Jones (LJ) potential. The solvent-solvent and solute-solute LJ parameters are determined by the fitting of the corresponding critical points via the equation of state. The solvent-solute parameters are determined using the standard Berthelot-Lorentz mixing rules. Critical parameters of the CO$_2$ are taken from NIST \cite{nist}, naproxen, ibuprofen, aspirin and dislunisal critical parameters are taken from the ref.\cite{garlapati2009temperature}, carbamazepine parameters -- from the ref. \cite{li2013new}, and the parameters of the thiadiazole derivative (1-[5-(3-chloro-4-methyl-phenylamino)-1,2,4-thiadiazol-3-yl]-propan-2-ol) were calculated, using group contribution methods \cite{tu1995group, lydersen1955estimation, klincewicz1984estimation}. The experimentally measured data of the vapor pressure temperature dependence are taken from the refs. \cite{perlovich2004naproxen,perlovich2004ibuprofen,perlovich2004aspirin,perlovich2003diflunisal,drozd2017novel,bui2014phycsico}. All values of the obtained parameters are represented in the Table \ref{table_supp_2}.

\begin{table}[h!]
\centering
\caption{Solute's critical temperature and pressure, molar volume, parameters of the solute-solute LJ interaction potential and coefficients of the empirical dependence of the solute's vapor pressure on temperature: $\ln(p^{sat})=A-B/T$.}
\begin{tabular}{l|c|l|l|c|c|c|c}
solute & $T_c$,K & $P_c$,bar & $v^s$, cm$^3\cdot$mol$^{-1}$ & $\sigma_{ss}$, $\si {\angstrom}$ & $\varepsilon_{ss},K $ & $A$ & $B$ \\
      \hline
naproxen	            &	807.00	&	24.52	&	179.0	&	7.356	&	580.418	&	39.7	&	15431	\\
ibuprofen	            &	749.70	&	23.0	&	182.1  	&	7.565	&	571.387	&	40.4	&	13927	\\
aspirin	                &	762.90	&	32.8	&	129 	&	6.553	&	548.700	&	38.2	&	13190	\\
carbamazepine	        &	786.83	&	25.71	&	180.48	&	7.180	&	565.911	&	32.7	&	13343	\\
diflunisal  	        &	869.80	&	32.11	&	125.5	&	6.894	&	625.585	&	36.4	&	14400	\\
thiadiazole derivative	&	895.83	&	23.67	&	202.16	&	7.708	&	644.306	&	33.2	&	13989	\\ \end{tabular}
\label{table_supp_2}
\end{table}

\begin{figure}[h!]
\center{\includegraphics[width=0.7\linewidth]{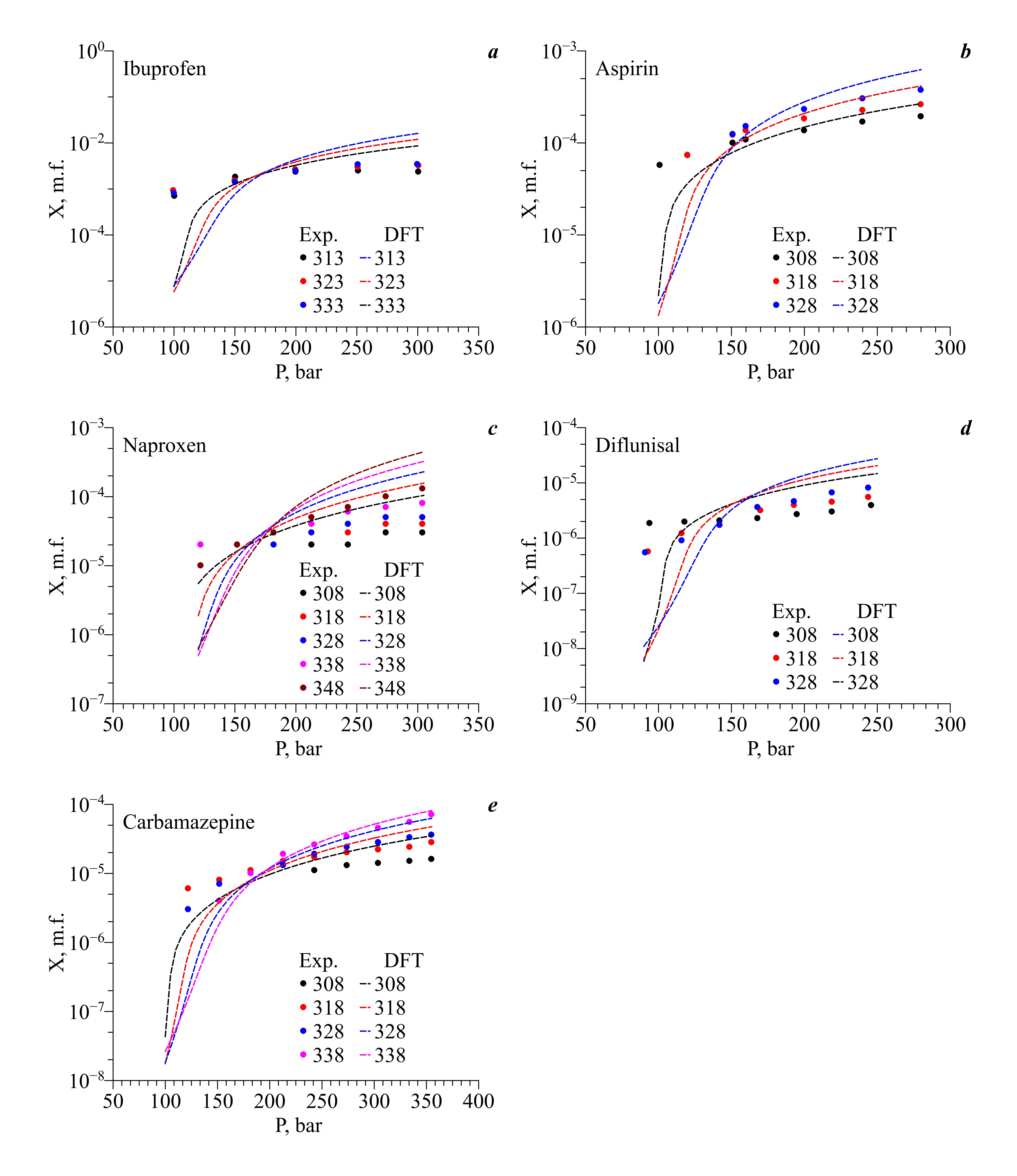}}
\caption{Comparison of the experimental solubility isotherms (colored circles) available in the literature with the ones obtained using cDFT-based approach (dashed lines).}
\label{Fig3supp}
\end{figure}

Validation of the solubility values correctness computed via the cDFT-based approach was provided by the comparison with the experimental solubility data, taken from the literature when available or obtained, basing on the experimental approach, described above. At the Fig. \ref{Fig3supp} we present the results of the isotherms comparison with the available in literature results. The corresponding experimental data are taken from: ibuprofen -- ref.\cite{kuznetsova2013solubility}, aspirin -- ref.\cite{ravipaty2008polar}, naproxen -- ref.\cite{garmroodi2004solubilities}, diflunisal -- ref.\cite{coimbra2008solubility} and carbamazepine -- ref.\cite{yamini2001solubilities}. One can see a reasonable divergence in the results for several compounds at low pressure values, but starting from the pressure around 150 bars the agreement is decent. Although the calculation then overestimate the solubility values at high pressures, we should note that these discrepancies do not exceed more than a half of the order magnitude. We find such agreement satisfactory as the originally proposed cDFT-based approach is supposed to be considered as a tool for the fast and sufficient estimation of the solute's solubility, rather than a technique used to obtain the high-accuracy solubility values.

\begin{figure}[h!]
\center{\includegraphics[width=0.7\linewidth]{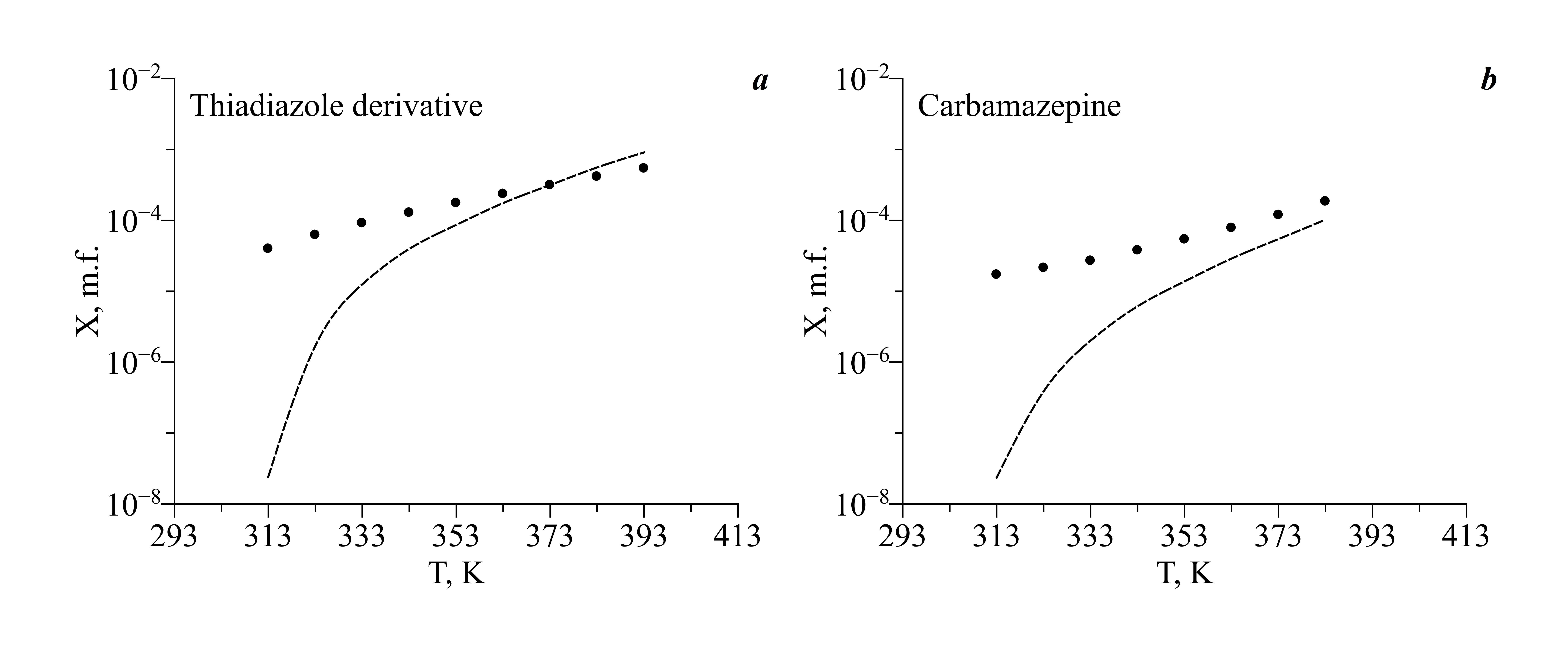}}
\caption{Comparison of the experimental solubility isochores (dots) with the ones obtained using cDFT-based approach (dashed line). Experimental data for carbamazepine were taken from the ref.\cite{kalikin2020carbamazepine}}
\label{Fig4supp}
\end{figure}

Experimentally measured thiadiazole derivative solubility values in comparison with the calculated ones are presented at the Fig. \ref{Fig4supp} alongside the same comparison for the carbamazepine, experimental data for which were taken from the ref.\cite{kalikin2020carbamazepine}. Once again the same trend can be observed but now regarding the temperature dependence, namely the underestimation at low temperatures and overestimation at high ones. Overall, starting from the 343 K the agreement is rather satisfactory.

\section*{Acknowledgments}

The research was supported by The Ministry of Science and Higher Education of the Russian Federation (grant no. RFMEFI61618$\times$0097). This research was supported through resources of supercomputer facilities provided by NRU HSE.

\bibliography{lit}

\begin{thebibliography}{10}
\expandafter\ifx\csname url\endcsname\relax
  \def\url#1{\texttt{#1}}\fi
\expandafter\ifx\csname urlprefix\endcsname\relax\def\urlprefix{URL }\fi
\expandafter\ifx\csname href\endcsname\relax
  \def\href#1#2{#2} \def\path#1{#1}\fi

\bibitem{chimowitz1986process}
E.~Chimowitz, K.~Pennisi, Process synthesis concepts for supercritical gas
  extraction in the crossover region, AIChE journal 32~(10) (1986) 1665--1676.

\bibitem{chimowitz1988analysis}
E.~Chimowitz, F.~Kelley, F.~Munoz, Analysis of retrograde behavior and the
  cross-over effect in supercritical fluids, Fluid phase equilibria 44~(1)
  (1988) 23--52.

\bibitem{foster1991significance}
N.~R. Foster, G.~S. Gurdial, J.~S. Yun, K.~K. Liong, K.~D. Tilly, S.~S. Ting,
  H.~Singh, J.~H. Lee, Significance of the crossover pressure in
  solid-supercritical fluid phase equilibria, Industrial \& engineering
  chemistry research 30~(8) (1991) 1955--1964.

\bibitem{esmaeilzadeh2008two}
F.~Esmaeilzadeh, Y.~Bozorgi, Two new thermodynamic constraints for a
  solid/supercritical fluid system, Chemical Engineering \& Technology:
  Industrial Chemistry-Plant Equipment-Process Engineering-Biotechnology
  31~(10) (2008) 1501--1509.

\bibitem{de2009solid}
S.~V. de~Melo, G.~M.~N. Costa, A.~Viana, F.~Pessoa, Solid pure component
  property effects on modeling upper crossover pressure for supercritical fluid
  process synthesis: A case study for the separation of annatto pigments using
  sc-co2, The Journal of Supercritical Fluids 49~(1) (2009) 1--8.

\bibitem{padrela2018supercritical}
L.~Padrela, M.~A. Rodrigues, A.~Duarte, A.~M. Dias, M.~E. Braga, H.~C.
  de~Sousa, Supercritical carbon dioxide-based technologies for the production
  of drug nanoparticles/nanocrystals--a comprehensive review, Advanced drug
  delivery reviews 131 (2018) 22--78.

\bibitem{rodrigues2018pharmaceutical}
M.~Rodrigues, B.~Baptista, J.~A. Lopes, M.~C. Sarragu{\c{c}}a, Pharmaceutical
  cocrystallization techniques. advances and challenges, International Journal
  of Pharmaceutics 547~(1-2) (2018) 404--420.

\bibitem{gurikov2018amorphization}
P.~Gurikov, I.~Smirnova, Amorphization of drugs by adsorptive precipitation
  from supercritical solutions: A review, The Journal of Supercritical Fluids
  132 (2018) 105--125.

\bibitem{cocero2009encapsulation}
M.~J. Cocero, {\'A}.~Mart{\'\i}n, F.~Mattea, S.~Varona, Encapsulation and
  co-precipitation processes with supercritical fluids: fundamentals and
  applications, The Journal of Supercritical Fluids 47~(3) (2009) 546--555.

\bibitem{kelley1990near}
F.~Kelley, E.~Chimowitz, Near-critical phenomena and resolution in
  supercritical fluid chromatography, AIChE journal 36~(8) (1990) 1163--1175.

\bibitem{esmaeilzadeh2005supercritical}
F.~Esmaeilzadeh, I.~Goodarznia, Supercritical extraction of phenanthrene in the
  crossover region, Journal of Chemical \& Engineering Data 50~(1) (2005)
  49--51.

\bibitem{wong1986solubilization}
J.~Wong, K.~Johnston, Solubilization of biomolecules in carbon dioxide based
  supercritical fluids, Biotechnology Progress 2~(1) (1986) 29--39.

\bibitem{swaid1985nir}
I.~Swaid, D.~Nickel, G.~Schneider, Nir—spectroscopic investigations on phase
  behaviour of low—volatile organic substances in supercritical carbon
  dioxide, Fluid Phase Equilibria 21~(1-2) (1985) 95--112.

\bibitem{kramer1988solubility}
A.~Kramer, G.~Thodos, Solubility of 1-hexadecanol and palmitic acid in
  supercritical carbon dioxide, Journal of Chemical and Engineering Data 33~(3)
  (1988) 230--234.

\bibitem{baek2004density}
J.-K. Baek, S.~Kim, G.-S. Lee, J.-J. Shim, Density correlation of solubility of
  ci disperse orange 30 dye in supercritical carbon dioxide, Korean Journal of
  Chemical Engineering 21~(1) (2004) 230--235.

\bibitem{pishnamazi2020measuring}
M.~Pishnamazi, S.~Zabihi, S.~Jamshidian, H.~Z. Hezaveh, A.~Z. Hezave,
  S.~Shirazian, Measuring solubility of a chemotherapy-anti cancer drug
  (busulfan) in supercritical carbon dioxide, Journal of Molecular Liquids 317
  (2020) 113954.

\bibitem{wang2015measurement}
T.-C. Wang, P.-C. Chang, Measurement and correlation for the solid solubility
  of antioxidants sodium l-ascorbate and sodium erythorbate monohydrate in
  supercritical carbon dioxide, Journal of Chemical \& Engineering Data 60~(3)
  (2015) 790--794.

\bibitem{knez2017solubility}
Z.~Knez, D.~C\"{o}r, M.~Knez~Hrn\v{c}i\v{c}, Solubility of solids in sub-and
  supercritical fluids: a review 2010--2017, Journal of Chemical \& Engineering
  Data 63~(4) (2017) 860--884.

\bibitem{gupta2006solubility}
R.~B. Gupta, J.-J. Shim, Solubility in supercritical carbon dioxide, CRC press,
  2006.

\bibitem{budkov2019possibility}
Y.~Budkov, A.~Kolesnikov, D.~Ivlev, N.~Kalikin, M.~Kiselev, Possibility of
  pressure crossover prediction by classical dft for sparingly dissolved
  compounds in scco2, Journal of Molecular Liquids 276 (2019) 801--805.

\bibitem{kalikin2020carbamazepine}
N.~Kalikin, M.~Kurskaya, D.~Ivlev, M.~Krestyaninov, R.~Oparin, A.~Kolesnikov,
  Y.~Budkov, A.~Idrissi, M.~Kiselev, Carbamazepine solubility in supercritical
  co2: a comprehensive study, Journal of Molecular Liquids (2020) 113104.

\bibitem{todd1990naproxen}
P.~A. Todd, S.~P. Clissold, Naproxen, Drugs 40~(1) (1990) 91--137.

\bibitem{rainsford2009ibuprofen}
K.~Rainsford, Ibuprofen: pharmacology, efficacy and safety,
  Inflammopharmacology 17~(6) (2009) 275--342.

\bibitem{cotton1985diflunisal}
M.~L. Cotton, R.~A. Hux, Diflunisal, in: Analytical profiles of drug
  substances, Vol.~14, Elsevier, 1985, pp. 491--526.

\bibitem{ballenger1978therapeutic}
J.~C. Ballenger, R.~M. Post, Therapeutic effects of carbamazepine in affective
  illness: a preliminary report., Communications in Psychopharmacology (1978).

\bibitem{awtry2000aspirin}
E.~H. Awtry, J.~Loscalzo, Aspirin, Circulation 101~(10) (2000) 1206--1218.

\bibitem{surov2015impact}
A.~Surov, C.~Bui, T.~Volkova, A.~Proshin, G.~Perlovich, The impact of
  structural modification of 1, 2, 4-thiadiazole derivatives on thermodynamics
  of solubility and hydration processes, Physical Chemistry Chemical Physics
  17~(32) (2015) 20889--20896.

\bibitem{gregorowicz2003solubility}
J.~Gregorowicz, Solubility of eicosane in supercritical ethane and ethylene,
  The Journal of supercritical fluids 26~(2) (2003) 95--113.

\bibitem{mendes1999solubility}
R.~L. Mendes, B.~P. Nobre, J.~P. Coelho, A.~F. Palavra, Solubility of
  $\beta$-carotene in supercritical carbon dioxide and ethane, The Journal of
  supercritical fluids 16~(2) (1999) 99--106.

\bibitem{schmitt1986solubility}
W.~J. Schmitt, R.~C. Reid, Solubility of monofunctional organic solids in
  chemically diverse supercritical fluids, Journal of Chemical and Engineering
  Data 31~(2) (1986) 204--212.

\bibitem{singh1993solubility}
H.~Singh, S.~J. Yun, S.~J. Macnaughton, D.~L. Tomasko, N.~R. Foster, Solubility
  of cholesterol in supercritical ethane and binary gas mixtures containing
  ethane, Industrial \& engineering chemistry research 32~(11) (1993)
  2841--2848.

\bibitem{kurnik1981solubility}
R.~T. Kurnik, S.~J. Holla, R.~C. Reid, Solubility of solids in supercritical
  carbon dioxide and ethylene, Journal of Chemical and Engineering Data 26~(1)
  (1981) 47--51.

\bibitem{tsai2014solubility}
C.-C. Tsai, H.-m. Lin, M.-J. Lee, Solubility of niflumic acid and celecoxib in
  supercritical carbon dioxide, The Journal of Supercritical Fluids 95 (2014)
  17--23.

\bibitem{miller1995determination}
D.~J. Miller, S.~B. Hawthorne, Determination of solubilities of organic solutes
  in supercritical co2 by online flame ionization detection, Analytical
  Chemistry 67~(2) (1995) 273--279.

\bibitem{miller1996solubility}
D.~J. Miller, S.~B. Hawthorne, A.~A. Clifford, S.~Zhu, Solubility of polycyclic
  aromatic hydrocarbons in supercritical carbon dioxide from 313 k to 523 k and
  pressures from 100 bar to 450 bar, Journal of Chemical \& Engineering Data
  41~(4) (1996) 779--786.

\bibitem{miller1997solubility}
D.~J. Miller, S.~B. Hawthorne, A.~A. Clifford, Solubility of chlorinated
  hydrocarbons in supercritical carbon dioxide from 313 to 413 k and at
  pressures from 150 to 450 bar, The Journal of Supercritical Fluids 10~(1)
  (1997) 57--63.

\bibitem{li2003solubility}
Q.~Li, Z.~Zhang, C.~Zhong, Y.~Liu, Q.~Zhou, Solubility of solid solutes in
  supercritical carbon dioxide with and without cosolvents, Fluid Phase
  Equilibria 207~(1-2) (2003) 183--192.

\bibitem{lou1997temperature}
X.~Lou, H.-G. Janssen, C.~A. Cramers, Temperature and pressure effects on
  solubility in supercritical carbon dioxide and retention in supercritical
  fluid chromatography, Journal of Chromatography A 785~(1-2) (1997) 57--64.

\bibitem{bartle1990measurement}
K.~D. Bartle, A.~A. Clifford, S.~A. Jafar, Measurement of solubility in
  supercritical fluids using chromatographic retention: the solubility of
  fluorene, phenanthrene, and pyrene in carbon dioxide, Journal of Chemical and
  Engineering Data 35~(3) (1990) 355--360.

\bibitem{pishnamazi2020using}
M.~Pishnamazi, S.~Zabihi, P.~Sarafzadeh, F.~Borousan, A.~Marjani, R.~Pelalak,
  S.~Shirazian, Using static method to measure tolmetin solubility at different
  pressures and temperatures in supercritical carbon dioxide, Scientific
  Reports 10~(1) (2020) 1--7.

\bibitem{shojaee2013experimental}
S.~A. Shojaee, H.~Rajaei, A.~Z. Hezave, M.~Lashkarbolooki, F.~Esmaeilzadeh,
  Experimental measurement and correlation for solubility of piroxicam (a
  non-steroidal anti-inflammatory drugs (nsaids)) in supercritical carbon
  dioxide, The Journal of Supercritical Fluids 80 (2013) 38--43.

\bibitem{pishnamazi2020thermodynamic}
M.~Pishnamazi, S.~Zabihi, S.~Jamshidian, F.~Borousan, A.~Z. Hezave,
  S.~Shirazian, Thermodynamic modelling and experimental validation of
  pharmaceutical solubility in supercritical solvent, Journal of Molecular
  Liquids 319 (2020) 114120.

\bibitem{zabihi2020loxoprofen}
S.~Zabihi, S.~H. Esmaeili-Faraj, F.~Borousan, A.~Z. Hezave, S.~Shirazian,
  Loxoprofen solubility in supercritical carbon dioxide: experimental and
  modeling approaches, Journal of Chemical \& Engineering Data 65~(9) (2020)
  4613--4620.

\bibitem{zabihi2021measuring}
S.~Zabihi, S.~Jamshidian, F.~Borousan, A.~Z. Hezave, M.~Pishnamazi, A.~Marjani,
  S.~Shirazian, Measuring salsalate solubility in supercritical carbon dioxide:
  experimental and thermodynamic modelling, The Journal of Chemical
  Thermodynamics 152 (2021) 106271.

\bibitem{sodeifian2019solubility}
G.~Sodeifian, F.~Razmimanesh, S.~A. Sajadian, Solubility measurement of a
  chemotherapeutic agent (imatinib mesylate) in supercritical carbon dioxide:
  Assessment of new empirical model, The Journal of Supercritical Fluids 146
  (2019) 89--99.

\bibitem{sodeifian2017determination}
G.~Sodeifian, S.~A. Sajadian, N.~S. Ardestani, Determination of solubility of
  aprepitant (an antiemetic drug for chemotherapy) in supercritical carbon
  dioxide: empirical and thermodynamic models, The Journal of Supercritical
  Fluids 128 (2017) 102--111.

\bibitem{oparin2016new}
R.~Oparin, E.~Vorobyev, M.~Kiselev, A new method for measuring the solubility
  of slightly soluble substances in supercritical carbon dioxide, Russian
  Journal of Physical Chemistry B 10~(7) (2016) 1108--1115.

\bibitem{oparin2014dynamic}
R.~D. Oparin, A.~Idrissi, M.~V. Fedorov, M.~G. Kiselev, Dynamic and static
  characteristics of drug dissolution in supercritical co2 by infrared
  spectroscopy: measurements of acetaminophen solubility in a wide range of
  state parameters, Journal of Chemical \& Engineering Data 59~(11) (2014)
  3517--3523.

\bibitem{oparin2019polymorphism}
R.~D. Oparin, Y.~A. Vaksler, M.~A. Krestyaninov, A.~Idrissi, S.~V. Shishkina,
  M.~G. Kiselev, Polymorphism and conformations of mefenamic acid in
  supercritical carbon dioxide, The Journal of Supercritical Fluids 152 (2019)
  104547.

\bibitem{wojdyr2010fityk}
M.~Wojdyr, Fityk: a general-purpose peak fitting program, Journal of Applied
  Crystallography 43~(5-1) (2010) 1126--1128.

\bibitem{immirzi1977prediction}
A.~Immirzi, B.~Perini, Prediction of density in organic crystals, Acta
  Crystallographica Section A: Crystal Physics, Diffraction, Theoretical and
  General Crystallography 33~(1) (1977) 216--218.

\bibitem{cao2008use}
X.~Cao, N.~Leyva, S.~R. Anderson, B.~C. Hancock, Use of prediction methods to
  estimate true density of active pharmaceutical ingredients, International
  journal of pharmaceutics 355~(1-2) (2008) 231--237.

\bibitem{nist}
E.~W. Lemmon, M.~O. McLinden, D.~G. Friend, Thermophysical properties of fluid
  systems, NIST chemistry WebBook, 1998.

\bibitem{garlapati2009temperature}
C.~Garlapati, G.~Madras, Temperature independent mixing rules to correlate the
  solubilities of antibiotics and anti-inflammatory drugs in scco2,
  Thermochimica Acta 496~(1-2) (2009) 54--58.

\bibitem{li2013new}
J.-h. Li, Z.~Huang, J.-l. Wei, L.~Xu, A new optimization method for parameter
  determination in modeling solid solubility in supercritical co2, Fluid Phase
  Equilibria 344 (2013) 117--124.

\bibitem{tu1995group}
C.-H. Tu, Group-contribution estimation of critical temperature with only
  chemical structure, Chemical engineering science 50~(22) (1995) 3515--3520.

\bibitem{lydersen1955estimation}
A.~Lydersen, Estimation of critical properties of organic compounds, Univ.
  Wisconsin Coll. Eng., Eng. Exp. Stn. Rep. 3 (1955).

\bibitem{klincewicz1984estimation}
K.~Klincewicz, R.~Reid, Estimation of critical properties with group
  contribution methods, AIChE Journal 30~(1) (1984) 137--142.

\bibitem{perlovich2004naproxen}
G.~L. Perlovich, S.~V. Kurkov, A.~N. Kinchin, A.~Bauer-Brandl, Thermodynamics
  of solutions iii: Comparison of the solvation of (+)-naproxen with other
  nsaids, European Journal of Pharmaceutics and Biopharmaceutics 57~(2) (2004)
  411--420.

\bibitem{perlovich2004ibuprofen}
G.~L. Perlovich, S.~V. Kurkov, L.~K. Hansen, A.~Bauer-Brandl, Thermodynamics of
  sublimation, crystal lattice energies, and crystal structures of racemates
  and enantiomers:(+)-and ($\pm$)-ibuprofen, Journal of pharmaceutical sciences
  93~(3) (2004) 654--666.

\bibitem{perlovich2004aspirin}
G.~L. Perlovich, S.~V. Kurkov, A.~N. Kinchin, A.~Bauer-Brandl, Solvation and
  hydration characteristics of ibuprofen and acetylsalicylic acid, Aaps
  Pharmsci 6~(1) (2004) 22--30.

\bibitem{perlovich2003diflunisal}
G.~L. Perlovich, S.~V. Kurkov, A.~Bauer-Brandl, Thermodynamics of solutions:
  Ii. flurbiprofen and diflunisal as models for studying solvation of drug
  substances, European journal of pharmaceutical sciences 19~(5) (2003)
  423--432.

\bibitem{drozd2017novel}
K.~V. Drozd, A.~N. Manin, A.~V. Churakov, G.~L. Perlovich, Novel drug--drug
  cocrystals of carbamazepine with para-aminosalicylic acid: Screening, crystal
  structures and comparative study of carbamazepine cocrystal formation
  thermodynamics, CrystEngComm 19~(30) (2017) 4273--4286.

\bibitem{bui2014phycsico}
C.~T. Bui, Physico-chemical properties of the crystals and solution of the
  1,2,4-thiadiazole based neuroprotective drug compounds (in russian), Ph.D.
  thesis, G.A. Krestov Institute of Solution Chemistry of the Russian Academy
  of Sciences (2014).

\bibitem{kuznetsova2013solubility}
I.~Kuznetsova, I.~Gilmutdinov, I.~Gilmutdinov, A.~Mukhamadiev, A.~Sabirzyanov,
  Solubility of ibuprofen in supercritical carbon dioxide, Russian Journal of
  Physical Chemistry B 7~(7) (2013) 814--819.

\bibitem{ravipaty2008polar}
S.~Ravipaty, K.~J. Koebke, D.~J. Chesney, Polar mixed-solid solute systems in
  supercritical carbon dioxide: entrainer effect and its influence on
  solubility and selectivity, Journal of Chemical \& Engineering Data 53~(2)
  (2008) 415--421.

\bibitem{garmroodi2004solubilities}
A.~Garmroodi, J.~Hassan, Y.~Yamini, Solubilities of the drugs benzocaine,
  metronidazole benzoate, and naproxen in supercritical carbon dioxide, Journal
  of Chemical \& Engineering Data 49~(3) (2004) 709--712.

\bibitem{coimbra2008solubility}
P.~Coimbra, D.~Fernandes, M.~H. Gil, H.~C.~d. Sousa, Solubility of diflunisal
  in supercritical carbon dioxide, Journal of Chemical \& Engineering Data
  53~(8) (2008) 1990--1995.

\bibitem{yamini2001solubilities}
Y.~Yamini, J.~Hassan, S.~Haghgo, Solubilities of some nitrogen-containing drugs
  in supercritical carbon dioxide, Journal of Chemical \& Engineering Data
  46~(2) (2001) 451--455.

\end{thebibliography}

\end{document}